\documentclass[onecolumn,pra,tightenlines,secnumarabic,preprintnumbers,floatfix,letterpaper,prl]{revtex4}
\usepackage{citesort,enumerate}
\usepackage{amsmath,amssymb} 
\usepackage{bm}              
\usepackage{amscd}           
\usepackage{epsfig}          
\usepackage{hhline,multirow} 
\usepackage{dcolumn}         
\newcommand{\half}{{1 \over 2}}
\newcommand{\third}{{1 \over 3}}

\begin{document}
\title{Role of the gluons in the color screening in a QCD plasma}
\author{G. Calucci}%
\email{giorgio@ts.infn.it}
\author{E. Cattaruzza}%
\email{ecattar@ts.infn.it}
\affiliation{%
Dipartimento di Fisica teorica dell'Universit\`a  \\
Strada Costiera 11, Miramare-Grignano, Trieste, I-34014\\
INFN, Sezione di Trieste, Italy}
\begin{abstract}
The color screening in a QCD plasma, that was studied in a formulation making
 evident similarities and differences with the electric case, is continued
 by taking into account the contributions of real gluons. The results, which 
 include a
 numerical analysis not previously performed, show a damping of the correlation
 function which, if not exponential, does not differ very much from that form.
 The role of the temperature, which affect both the population and the dynamics
 of the quark-gluon system, is found to be relevant.
\end{abstract}
 \maketitle
 \section {1. Introduction }
   The analysis of the screening in $q \bar q-$plasma performed keeping the
 analogy with the classical treatment of an electric plasma \cite{Calucci} gave as result a
 slight difference in the screening behavior with distance, with respect to the
 more standard, exponential decay \cite{Birse, Philipsen, Kogut}. Now this analysis is pushed further by
 introducing the effect of the real gluons which are certainly present in such
 situations. Once the relevance of the gluons in the quark-quark correlation 
 functions
 is taken into account one in lead, unavoidably, to consider also the
 quark-gluon correlations and this is performed in the present paper. While a 
 definite quark population is taken as a initial condition of the problem, the 
 gluon population is considered of thermal origin and is given therefore in 
 terms of a
 Bose-Einstein density, together with this population also the creation of 
 $q \bar q-$pair is considered, with a corresponding Fermi-Dirac density.
  Being interested in the large distance effects, that part of the interaction of real
 gluons with quark is considered, which correspond to a $t-$channel exchange of a
 virtual gluon, so that it has, suitably redefining color and spin factors, the
 same form of quark-quark interaction.\par
 The result is a general behavior of the correlation function not very different
 from that previously found in a pure $q \bar q-$plasma, since the analytical
 form is however awkward a numerical investigation has been performed for
 both situations without gluons and with gluons. The behavior differs little
 from the exponential, in both case the possibility of small oscillations is
 found, but only when the correlation function is already very small.\par
 A discussion of the range of values for which the model can be meaningful and
 also a comparison with some different treatments and results, in particular 
 with the formulation using the termal Green functions are briefly presented \cite{Le-Bellac}.
\par
 Three short Appendices contains some details of the needed calculations.
\section{2.Study of the Correlation Functions}
\subsection{2.1 General form of the equations}
As it has been explained in the Introduction the basic equation for the
correlation functions have the same general features as for the pure $q\bar q$
populations, but we must distinguish two possible correlation, the
two-fermion case and the fermion-boson case.
In principle we should look also at the two-boson correlation function, since
however our main interest is in the two-fermion correlations while keeping into
account the boson-fermion correlation, that enter directly in the equation for
the two-fermion function we neglect the effect of the pure bosonic correlation:
in so doing we are forced to bound our discussion to situation where the fermion
density is dominant, having in mind the nucleus-nucleus collision there are
certainly situations, of not too large temperature, where this condition is
satisfied.\par
The ways in which the usual determination of the shielding has been adapted to
the case of non commuting charges have been presented in the provios paper and
will not be discussed here, only the integrodifferential equation for the two 
body correlation is reproduced

  \begin{align} 
   &{{ \partial C_{\beta}(q_1,q_2)}\over{\partial r_{1,v}}}= 
   -\int^{\beta}_o d\tau \bigg[
    {{\partial u(q_1,q_2)}\over{\partial r_{1,v}}}+\cr
   &{1\over {3V}}\sum_{l\neq 1,2}\int d^3 q_l\Big[
   {{\partial u(q_1,q_l)}\over{\partial r_{1,v}}}
   C_{\beta-\tau}(q_l,q_2)+
   C_{\tau}(q_l,q_2){{\partial u(q_1,q_l)}\over{\partial r_{1,v}}}
   \Big]\bigg] \label{2-Correlation}\end{align}
   
 where the variables $q$ embody both the space variables $r$ and the color
 indices and the integration over the auxiliary variable $\tau$ takes care of 
 the non commutativity of the charges.

   Now a further derivative with respect to $r_1$ is calculated, then 
   one performs the Fourier transform with respect to the space variables and the
   Laplace transform with respect to the inverse temperature with the
   result:
   \begin{equation}
   -k^2 \check C(s;k)={{4\pi\alpha T}\over {s^2}}+{{4\pi\alpha}\over {3Vs}}
   \sum \Big[T \check C(s;k)+\check C(s;k)T  \Big] \label{2-Correlation-transform}
   \end{equation}
   The check (as $\check C$) means both the Fourier and
   the Laplace transform; the variable $k$ is the Fourier-conjugated of $r$ and
   $s$ is the Laplace-conjugated of the inverse temperature $\beta$, $4\pi\alpha=g^2$.
   The factors $\check C(s;k)$ and $T$ are
   matrices so their order is relevant. In order to proceed is is necessary to
   specify the colour structure of the quarks and of the gluons, so the indices will now
   must now be displayed and the precise meaning of $T$ and $\check C$ is
   specified. It is not convenient to work with two set of indices, for quark
   and gluon respectively, so the color charge of the gluon is represented by a
   traceless tensor with triplet indices, like $A^c_d$ with $A^n_n=0$.\par
   The term $T$ is specified in this way: the interactions
   $qq$ and $\bar q\bar q$ are given by [App.A]
 $$I_{a,c}^{b,d}=
  \frac{1}{2}\left[\delta_a^d\,\delta_c^b-
  \frac{1}{3}\,\delta_a^b\,\delta_c^b\right]$$
   the interaction $q\bar q$ is given by $-I$; the interaction
   $qg$ is given by 
  $$J_{a,cf}^{b,dg}=\frac{1}{2\,i}
  \left[\delta_a^d\,\delta_c^b\,\delta_f^g-
  \,\delta_a^g\,\delta_c^d\,\delta_f^b\right]$$
  the interaction
   $\bar qg$ is given by $-J$.\par
\begin{figure}[t]
\begin{center}
\vskip-1.cm
\includegraphics[scale=0.60,angle=0]{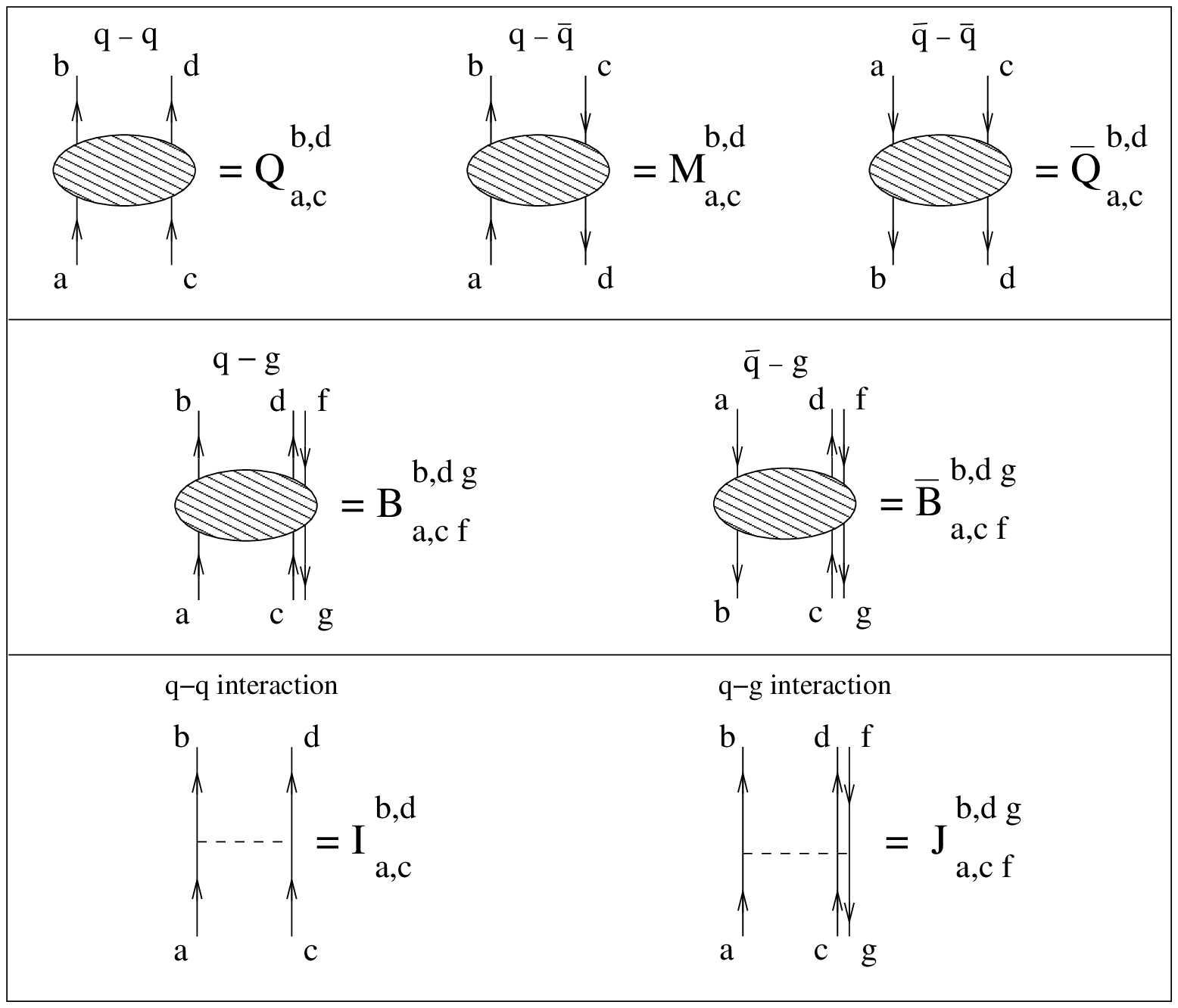}
\caption{Rules for a graphical interpretation of eq.(\ref{color-eq}): up and down arrows indicate respectively quarks and antiquarks while the gluons are represented in the form of a $q \bar q$ pair(up-down arrow).}
\label{Rules-equation}
\includegraphics[scale=0.60,angle=0]{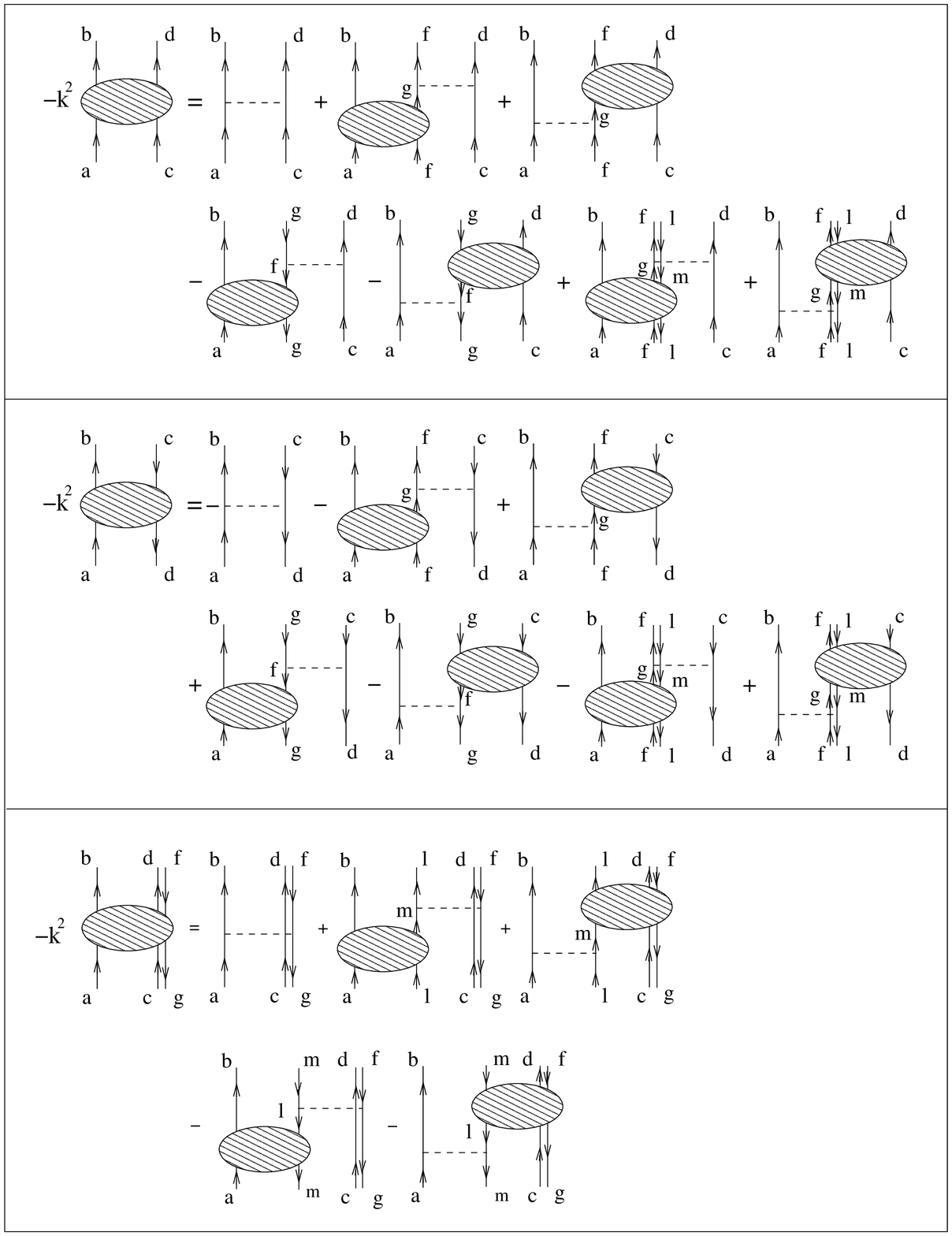}
\caption{Graphical representation of equations eq.(\ref{color-eq})}
\label{Color-eq-graph}
\end{center}
\end{figure}

   The term $\check C$ is indicated with different symbols following its quark
   and gluon content and is decomposed according to the color structure:
    for $qq$ and $\bar q\bar q$ it is called respectively $Q$ and $\bar Q$ and
    decomposed in terms
   of {\it triplet} and {\it sextet} for $q\bar q$ it is called $M$ and
   decomposed into
   {\it singlet} and {\it octet} and for $qg$ or $\bar qg$ it is called $B$ or 
   $\bar B$ and decomposed into {\it triplet,} {\it sextet} and {\it 15-plet} 
   by means of the following projectors:
\begin{align}
{}^3\Pi_{a,c}^{b,d}&=\frac{1}{2}\left[\delta_a^b\,\delta_c^d-\,\delta_a^d\,\delta_c^b\right]&
{}^6\Pi_{a,c}^{b,d}&=\frac{1}{2}\left[\delta_a^b\,\delta_c^d+\,\delta_a^d\,\delta_c^b\right]\nonumber\\
{}^1\Pi_{a,c}^{b,d}&=\frac{1}{3}\,\delta_a^d\,\delta_a^d&
{}^8\Pi_{a,c}^{b,d}&=\left[\delta_a^b\,\delta_c^d-\frac{1}{3}\,\delta_a^d\,\delta_c^b\right]&
\end{align}
 \begin{align} 
{}^3P_{a,cf}^{b,dg}& =\frac{3}{8}\,\delta_{a}^{g}\,\delta_{c}^{d}\,\delta_{f}^{b}-
\frac{1}{8}\,\left[\delta_{a}^{d}\,\delta_{c}^{g}\,\delta_{f}^{b}+\delta_{a}^{g}\,\delta_{c}^{b}\,
\delta_{f}^{d}\right]+\frac{1}{24}\,\delta_{a}^{b}\,\delta_{c}^{g}\,\delta_{f}^{d}\nonumber\\
{}^6P_{a,cf}^{b,dg}& = \frac{1}{2}\,\left[\delta_{a}^{b}\,\delta_{c}^{d}-\delta_{a}^{d}\,\delta_{c}^{b}\right]\,\delta_{f}^{g}-\frac{1}{4}\,\left[\delta_{a}^{g}\,\delta_{c}^{d}\,\delta_{f}^{b}
+\delta_{a}^{b}\,\delta_{c}^{g}\,\delta_{f}^{d}
-\delta_{a}^{g}\,\delta_{c}^{b}\,\delta_{f}^{d}
-\delta_{a}^{d}\,\delta_{c}^{g}\,\delta_{f}^{b}\right]\nonumber\\
{}^{15}P_{a,cf}^{b,dg}& = \frac{1}{2}\,\left[\delta_{a}^{b}\,\delta_{c}^{d}+\delta_{a}^{d}\,\delta_{c}^{b}\right]\,\delta_{f}^{g}-\frac{1}{8}\,\left[\delta_{a}^{g}\,\delta_{c}^{d}\,\delta_{f}^{b}
+\delta_{a}^{b}\,\delta_{c}^{g}\,\delta_{f}^{d}
+\delta_{a}^{g}\,\delta_{c}^{b}\,\delta_{f}^{d}
+\delta_{a}^{d}\,\delta_{c}^{g}\,\delta_{f}^{b}\right]
\end{align}
The correct normalization of the projector is easily verified
\begin{gather}
{}^1\Pi_{f,g}^{f,g}=1\quad {}^8\Pi_{f,g}^{f,g}=8\quad {}^3\Pi_{f,g}^{f,g}=3\quad {}^6\Pi_{f,g}^{f,g}=6\nonumber\\
{}^3P_{f,g,h}^{f,g,h}=3\quad {}^6P_{f,g,h}^{f,g,h}=6\quad {}^{15}P_{f,g,h}^{f,g,h}=15\nonumber
\end{gather}
For the different amplitudes, the equations corresponding to the general form of
eq.(\ref{2-Correlation-transform}) are

\begin{align}
\begin{split}
-k^2\,M_{a,c}^{b,d}& =-\frac{4\,\pi\,\alpha}{s^2}\,I_{a,c}^{b,d}+\frac{4\,\pi\,\alpha}{s}\,\left[\left(Q_{a,f}^{b,g}\,(-I_{g,c}^{f,d})+I_{a,f}^{b,g}\,M_{g,c}^{f,d})\right)\,\rho\right.\\
& + \left.\left(M_{a,f}^{b,g}\,I_{g,c}^{f,d}+(-I_{a,f}^{b,g})\,\bar Q_{g,c}^{f,d})\right)\,\bar \rho+\left(B_{a,fm}^{b,gl}\,(-J_{c,gl}^{d,fm})+J_{a,fm}^{b,gl}\,\bar B_{c,gl}^{d,fm}\right)\,\gamma\right]\\
-k^2\,Q_{a,c}^{b,d}& =\frac{4\,\pi\,\alpha}{s^2}\,I_{a,c}^{b,d}+\frac{4\,\pi\,\alpha}{s}\,\left[\left(Q_{a,f}^{b,g}\,I_{g,c}^{f,d}+I_{a,f}^{b,g}\,Q_{g,c}^{f,d})\right)\,\rho\right.\\
& + \left.\left(M_{a,f}^{b,g}\,(-I_{g,c}^{f,d})+(-I_{a,f}^{b,g})\,\bar M_{g,c}^{f,d})\right)\,\bar \rho+\left(B_{a,fm}^{b,gl}\,J_{c,gl}^{d,fm}+J_{a,fm}^{b,gl}\, B_{c,gl}^{d,fm}\right)\,\gamma\right]\\
-k^2\,B_{a,cf}^{b,dg}& =\frac{4\,\pi\,\alpha}{s^2}\,J_{a,cf}^{b,dg}+\frac{4\,\pi\,\alpha}{s}\,\left[\left(Q_{a,l}^{b,m}\,J_{m,cf}^{l,dg}+I_{a,l}^{b,m}\,B_{m,cf}^{l,dg}\right)\,\rho\right.\\
& + \left.\left(M_{a,l}^{b,m}\,(-J_{m,cf}^{l,dg})+(-I_{a,l}^{b,m})\,\bar B_{m,cf}^{l,dg})\right)\,\bar \rho \right]\label{color-eq}
\end{split}
\end{align}
The coefficients $\rho,\;\bar\rho,\;\gamma$ give the densities of quarks,
antiquarks ans gluons; the equations for $\bar Q$ and $\bar B$ can be obtained 
by interchange 
$Q\Leftrightarrow \bar Q$, $\rho \Leftrightarrow \bar \rho$ and 
$B \Leftrightarrow -\bar B$ in the equations for $Q$ and $B$. 
It is possible to give a graphical representation to eq.(\ref{color-eq}), see FIG.\ref{Color-eq-graph}, through
the rules set in FIG.\ref{Rules-equation}; it is consistent with the colour structure to use lower indices for incoming quarks and outgoing antiquarks and upper indices for incoming antiquarks and outgoing quarks. Since the gluon colour structure is given in the spinorial version [App.A], the gluon is represented in the form of a $q\bar q$ pair, where up-arrow and down-arrow are  used respectively for quarks and antiquarks.    
\subsection{2.2 Solution of the equations in the conjugate variables}
We recall that the interaction term $J$ is pure imaginary; in the last of eq.(\ref{color-eq}) it can be seen that 
$B$ depends linearly on J, in the equations for $M$ and $Q$, however, $B$ and
$J$ appear always together and always multiplied by $\gamma$, so it is useful
to perform the substitutions $J=-iJ'\,;B=-iB'$; in this way in the last equation we get a
trivial factor $-i$, in the other two we get a change of sign in the terms that
multiplies $\gamma $. In the whole set of eq (5) this amounts to
eliminate the "prime" in $J$ and $B$ and to change the sign $\gamma \to- \gamma$.
When we perform the decomposition
\begin{gather}
M={}^1\Pi\,F_1+{}^8\Pi\,F_8,\quad Q={}^3\Pi\,F_3+{}^6\Pi\,F_6\quad \bar Q={}^3\Pi\,\bar F_3+{}^6\Pi\,\bar F_6\nonumber\\
B={}^3P\,B_3+{}^6P\,B_6+{}^{15}P\,B_{15}\quad \bar B={}^3P\,\bar B_3+{}^6P\,\bar B_6+{}^{15}P\,\bar B_{15}\nonumber
\end{gather}
 we find contracting the interaction term $J$ with the projectors $P$ that
 the RHS of  equations (\ref{color-eq}) for the ($qq,\,q\bar q,\bar q\bar q$)
 systems contain only the $I_{a,c}^{b,d}$ tensor; in fact in any case an octet is exchanged in the t-channel, in this way the relations  
\begin{gather}
F_8=-\frac{1}{8}\,F_1,\quad F_6= -\half\,F_3,\quad 
\bar F_6= -\frac{1}{2}\,\bar F_3 \nonumber 
\end{gather}
hold as in the case without gluons and we get the equations:
\begin{align}
&k^2\,F_1-\frac{16\,\pi\,\alpha}{3\,s^2}+\frac{4\,\pi\,\alpha}{s}\left[F_3\,\rho+\bar F_3\,\bar \rho+
F_1\,\left(\frac{\rho+\bar \rho}{2}\right)+2\left(\bar B_3+\bar B_6-B_3-B_6\right)\,\gamma\right]=0\nonumber\\
&k^2\,F_3-\frac{8\,\pi\,\alpha}{3\,s^2}+\frac{4\,\pi\,\alpha}{s}\left[F_3\,\rho+
F_1\,\left(\frac{\bar \rho}{2}\right)-2\left(B_3+B_6\right)\,\gamma\right]=0\label{F1-F3-F3bar}
\\
&k^2\,\bar F_3-\frac{8\,\pi\,\alpha}{3\,s^2}+\frac{4\,\pi\,\alpha}{s}\left[\bar F_3\,\bar \rho+
F_1\,\left(\frac{\rho}{2}\right)+2\left(\bar B_3+\bar B_6\right)\,\gamma\right]=0,\nonumber
\end{align}
where $B_{15}$ and $\bar B_{15}$ have been substituted by means of the following relations:  
\begin{align}
B_{x,ac}^{x,bd}&=\bar B_{x,ac}^{x,bd}=0 \Rightarrow \,\quad 3\,B_3+6\,B_6+15\,B_{15}=0,\,\quad 3\,\bar B_3+6\bar B_6+15\,\bar B_{15}=0.\nonumber
\end{align}
The system (\ref{F1-F3-F3bar}) yields immediately $F_1=F_3+\bar F_3$ and is reduced to a 
two-equation system:
\begin{align}
&F_3\left[k^2+\frac{4\,\pi\,\alpha}{s}\,\left(\rho+\frac{\bar \rho}{2}\right)\right]+\bar F_3\,\left[\frac{2\,\pi\,\alpha\,\bar \rho}{s}\right]-\frac{8\,\pi\,\alpha}{3\,s^2}-\frac{8\,\pi\,\alpha}{s}\,\gamma\,(B_3+B_6)=0\nonumber\\
&F_3\,\left[\frac{2\,\pi\,\alpha\, \rho}{2\,s}\right]+ \bar F_3\left[k^2+\frac{4\,\pi\,\alpha}{s}\,\left(\bar \rho+\frac{ \rho}{2}\right)\right]-\frac{8\,\pi\,\alpha}{3\,s^2}+\frac{8\,\pi\,\alpha}{s}\,\gamma\,(\bar B_3+\bar B_6)=0.\nonumber
\end{align}
From the previous linear system it results that $F_3$ and $\bar F_3$ can be expressed in terms of $(B_3+B_6)$ and $(\bar B_3+\bar B_6)$, as follows: 
\begin{align}
&F_3=F_3^{(0)}+f_3^{(1)}\,(B_3+B_6)+f_3^{(2)}\,(\bar B_3+\bar B_6)
&\bar F_3=\bar F_3^{(0)}+\bar f_3^{(1)}\,(B_3+B_6)+\bar f_3^{(2)}\,(\bar B_3+\bar B_6)\label{eq:F3-F3BAR}
\end{align} 
where
\begin{align}
&F_3^{(0)}=\bar F_3^{(0)}=\frac{8\,\pi\,\alpha}{3\,s\,(k^2\,s+4\,\pi\,\alpha\,n)}\nonumber\\
&f_3^{(1)}=\frac{8\,\pi\,\alpha\,\gamma}{k^2\,s+4\,\pi\,\alpha\,n}+
\frac{16\,\pi^2\,\alpha^2\,\gamma\,\bar \rho}{(k^2\,s+4\,\pi\,\alpha\,n)\,(k^2\,s+2\,\pi\,\alpha\,n)}\quad
f_3^{(2)}=\frac{16\,\pi^2\,\alpha^2\,\gamma\,\bar \rho}{(k^2\,s+4\,\pi\,\alpha\,n)\,(k^2\,s+2\,\pi\,\alpha\,n)}\nonumber\\
&\bar f_3^{(1)}=-\frac{16\,\pi^2\,\alpha^2\,\gamma\, \rho}{(k^2\,s+4\,\pi\,\alpha\,n)\,(k^2\,s+2\,\pi\,\alpha\,n)}\quad
 \bar f_3^{(2)}=-\frac{8\,\pi\,\alpha\,\gamma}{k^2\,s+4\,\pi\,\alpha\,n}-
\frac{16\,\pi^2\,\alpha^2\,\gamma\,\rho}{(k^2\,s+4\,\pi\,\alpha\,n)\,(k^2\,s+2\,\pi\,\alpha\,n)}\nonumber
\end{align}
and $n=\rho+\bar \rho$. Furthermore we get
\begin{align}
&F_1=2\,F_3^{(0)}+f_1^{(1)}\,(B_3+B_6)+f_1^{(2)}\,(\bar B_3+\bar B_6)\nonumber
\end{align} 
\begin{align}
&f_1^{(1)}=\frac{8\,\pi\,\alpha\,\gamma\,(k^2\,s+4\,\pi\,\alpha\,\bar \rho)}{(k^2\,s+4\,\pi\,\alpha\,n)\,(k^2\,s+2\,\pi\,\alpha\,n)}\quad
&f_1^{(2)}=-\frac{8\,\pi\,\alpha\,\gamma\,(k^2\,s+4\,\pi\,\alpha\,\rho)}{(k^2\,s+4\,\pi\,\alpha\,n)\,(k^2\,s+2\,\pi\,\alpha\,n)}
\nonumber\end{align} 
Projecting the equations satisfied by $B$ and $\bar B$ with ${}^JP$ the following equations are obtained:
\begin{align}
&k^2\,B_3-\frac{6\,\pi\,\alpha}{s^2}+\frac{2\,\pi\,\alpha}{s}\,\left[\frac{9}{4}\,F_3\,\rho+\frac{9}{8}\,F_1\,\bar \rho+(B_3\,\rho-\bar B_3\,\bar \rho)\right]=0\label{eq:B3}\\
&k^2\,B_6-\frac{2\,\pi\,\alpha}{s^2}+\frac{2\,\pi\alpha}{s}\,\left[\frac{3}{4}\,F_3\,\rho+\frac{3}{8}\,F_1\,\bar \rho+(B_6\,\rho-\bar B_6\,\bar \rho)\right]=0\label{eq:B6}
\end{align}
The equation for $\bar B_3,\,\bar B_6$ can obtained by the interchange $B_i \Leftrightarrow  -\bar B_i$ for $i=3,6$, $\rho \Leftrightarrow \bar  \rho$ and $F_3 \Leftrightarrow \bar F_3$ in the equations for $ B_3,\, B_6$.  
After some tedious calculations, the following two equations for $(B_3+B_6)$ and $(\bar B_3+\bar B_6)$ variable are obtained:
\begin{align}
&\left[k^2+\frac{2\,\pi\,\alpha}{s}\,\left(3\,f_3^{(1)}\,\rho+\frac{3}{2}\,f_1^{(1)}\,\bar \rho+\rho\right)\right]\,(B_3+B_6)+\left[\frac{2\,\pi\,\alpha}{s}\,\left(3\,f_3^{(2)}\,\rho+\frac{3}{2}\,f_1^{(2)}\,\bar \rho-\bar \rho\right)\right]\,(\bar B_3+\bar B_6)\nonumber \\\quad&-\frac{8\,\pi\,\alpha}{s^2}+\frac{6\,\pi\,\alpha}{s}\,F_3^{(0)}\,(\rho+\bar \rho)=0\nonumber\\
&\left[\frac{2\,\pi\,\alpha}{s}\,\left(3\,\bar f_3^{(1)}\,\bar \rho+\frac{3}{2}\,f_1^{(1)}\, \rho+\rho\right)\right]\,(B_3+B_6)+\left[-k^2+\frac{2\,\pi\,\alpha}{s}\,\left(3\,\bar f_3^{(2)}\,\bar \rho+\frac{3}{2}\,f_1^{(2)}\, \rho-\bar \rho\right)\right]\,(\bar B_3+\bar B_6)\nonumber \\\quad&-\frac{8\,\pi\,\alpha}{s^2}+\frac{6\,\pi\,\alpha}{2\,s}\,F_3^{(0)}\,(\rho+\bar \rho)=0\nonumber
\end{align}
Solving the previous linear system we have:
\begin{align}
B_3+B_6& = \frac{8\,\pi\,\alpha\,(k^2\,s+2\,\pi\,\alpha\,n)}{s\,(k^4\,s^2+6\,k^2
\,s\,\pi\,\alpha\,n+48\,\pi^2\,\alpha^2\,\gamma\,n+8\,\pi^2\,\alpha^2\,n^2)}\nonumber\\
\bar B_3+\bar B_6& =-\frac{8\,\pi\,\alpha\,(k^2\,s+2\,\pi\,\alpha\,n)}{s\,(k^4\,s^2+6\,k^2
\,s\,\pi\,\alpha\,n+48\,\pi^2\,\alpha^2\,\gamma\,n+8\,\pi^2\,\alpha^2\,n^2)}= -(B_3+B_6) \nonumber
\end{align}
Substituting the previous expressions in eq.(\ref{eq:F3-F3BAR}) we have:
\begin{align}
F_3=  \bar F_3& = F_3^{(0)}+\frac{32\,\pi^2\,\alpha^2\,\gamma}{s\,(k^4\,s^2+6\,k^2
\,s\,\pi\,\alpha\,n+48\,\pi^2\,\alpha^2\,\gamma\,n+8\,\pi^2\,\alpha^2\,n^2)}\nonumber\\
&+\frac{32\,\pi^2\,\alpha^2\,\gamma}{(k^4\,s^2+6\,k^2
\,s\,\pi\,\alpha\,n+48\,\pi^2\,\alpha^2\,\gamma\,n+8\,\pi^2\,\alpha^2\,n^2)\,(k^2\,s+4\,\pi\,\alpha\,n)}
\label{F3-solution}
\end{align}
 From eq.(\ref{eq:B3},\ref{eq:B6},\ref{eq:B15},\ref{F3-solution}) it can be easily verified that: 
\begin{align}
B_3=-\bar B_3,\quad B_6=-\bar B_6,\quad B_{15}=-\bar B_{15}
\end{align}
\subsection {2.3 Correlations in real space}
\begin{figure}[t]
\begin{center}
\includegraphics[scale=0.50,angle=270]{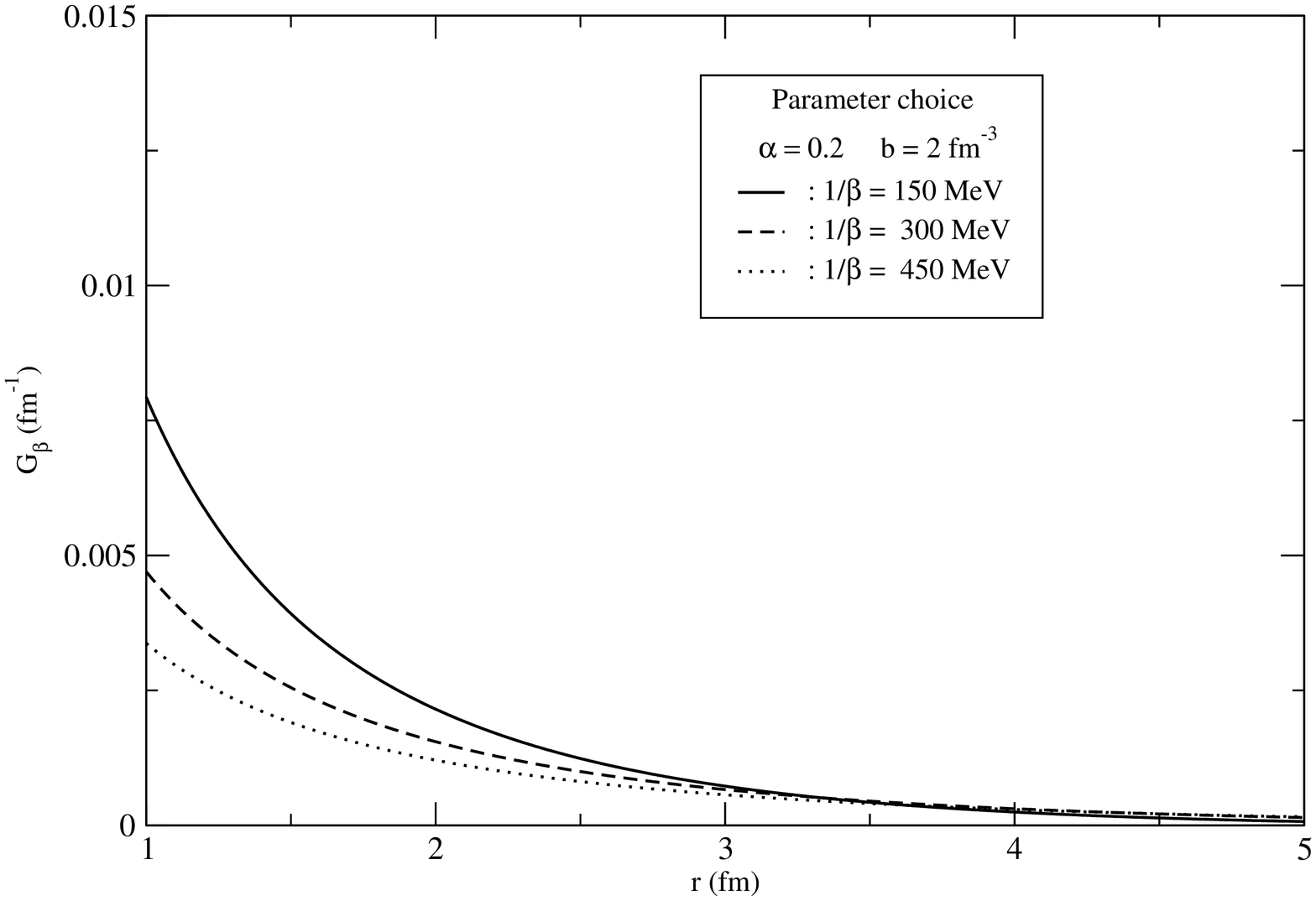}
\vskip-1.cm
\caption{Correlation function $G_{\beta}(r)$ in absence of gluons for different choices of temperatures, at fixed coupling constant $\alpha=0.2$ and initial quark density $b=2\,fm^{-3}$.}
\label{qq-correlation}
\includegraphics[scale=0.50,angle=270]{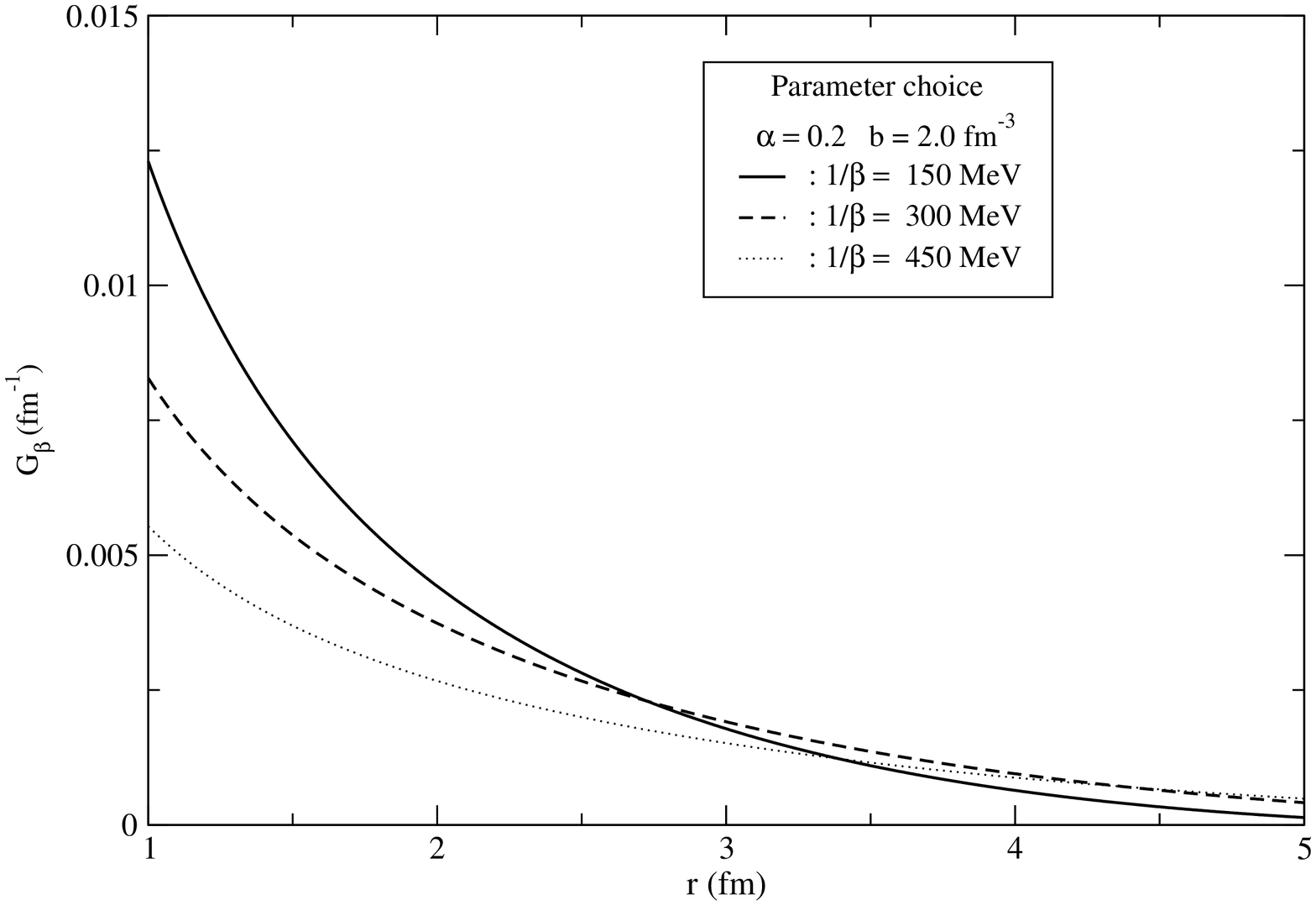}
\vskip-1.cm
\caption{Correlation function $G_{\beta}(r)$ in presence of gluons for different choices of temperatures, at fixed coupling constant $\alpha=0.2$ and initial quark density $b=2\,fm^{-3}$.}
\label{qq-gg-correlation}
\end{center}
\end{figure}

All the $q\bar q-$correlations may be expressed in term of $F_3$, so we  
evaluate its Laplace antitrasform with the following result:
\begin{align}
\hat G_{\beta}(k^2)= \frac{2}{3\,n}\,
\left[1+\frac{6\,\gamma / n}{1+6\,\gamma /n}\right]\,
\left[1-e^{-\frac{3\,\pi\,n\,\alpha\,\beta}{k^2}}
\left(\cos\left(\frac{3\,\pi\,n\,\alpha\,\beta\,y }{k^2}\right)-
\frac{1-12\,\gamma /n}{3\,y\,(1+12\,\gamma /n)}\,
\sin\left(\frac{3\,\pi\,n\,\alpha\,\beta\,y}{k^2}\right)\right)\right]
\label{F3laplace}
\end{align}
where $y =\frac{1}{3}\,\sqrt{48\,\gamma /n-1}\,$. 
 The inversion of the Fourier transform leads to an expression that is very 
 little transparent, it is reported for completeness in [App.C]; the correlation function in real space has been computed 
 numerically, by means of standard integration subroutines \cite{Vegas}, starting form the following integral
 \begin{equation}  
 	G_{\beta}(r^2)=\frac{1}{2\,\pi\,r^2}\int_0^{\infty} dk\,k\,\sin(k\,r)\,\hat G_{\beta}(k^2); 
 \end{equation}
In FIG.\ref{qq-correlation} and FIG.\ref{qq-gg-correlation} the correlation function without and with gluons for different choices of temperature, at fixed coupling constant $\alpha = 0.2$ and initial quark density $b=2\,fm^{-3}$, is shown.  
In both the cases it can be noted that the damping of the correlation, if not exponential, differs very little from that behavior; there is also a long distance $r$ region (with $r\sim 8\,fm$ at $1/\beta = 350\, MeV$), where the correlation begin to oscillate, but the amplitude of these oscillations is very small, being present when the correlation has been already much damped. 
The previous plots give a good overall description of the behavior of the correlation function, but do not show immediately how these functions differ from a Yukawa shape $e^{-\mu\,r}/r$; in order to give a more complete description of this property in Fig.\ref{qq-gg-correlation-lenght} the expression 
$-\ln(r\,G_{\beta}(r))$ is plotted. If the Yukawa shape was exact, we would find a straight line; these plots confirm (obviously) that the gluons make the damping weaker, moreover the correlation function becomes slightly farther from a Yukawa shape.  
It can be seen that the effect associated to the gluon presence is to produce, at fixed temperature, a perceptible increase in the value of the correlation and consequently a displacement towards higher $r$ values of the region where the correlation function begin to oscillate. For a detailed analysis
of the values of energies, quark, anti-quark and gluon densities, we refer to the next section and [App.B].   
\maketitle
\section{3. Conclusions}
The inclusion of the gluon in the effect of mutual screening in a $q\bar
q-$plasma refines a previous analysis given in term of a pure quark-antiquark
population, and it confirms the results. In particular the correlation length 
and the shape of the damping is still the same for $qq$ and $q\bar q$, so that 
it must reflect in the same way in the meson and baryon production.\par
An item that requires some condideration is the comparison with other ways of
dealing with the same phenomenon. The treatment given in term of strong coupling
on the lattice is difficult to compare with  \cite{Birse, Philipsen, Kogut} because it is very different
since the beginning, more similar are the treaments in terms of thermal Green
function as in \cite{Le-Bellac}. In that case one extracts the contribution of
the pole of the propagator in momentum space so that a precise Yukawa-like decay 
of the correlation function is certainly produced. The treatment here presented
is purely static, so it seems that it lacks of some effects that are present in
the other treatment, does it contains also someting more? Perhaps what is more
is seen looking at the structure of the coupled equation (\ref{color-eq}): the
interaction of one particle takes place with other particles which are already
correlated, so that at least at the level of two-fermion distribution a self 
consistent treatment of the correlations is performed.\par
\begin{figure}[t]
\begin{center}
\includegraphics[scale=0.70,angle=270]{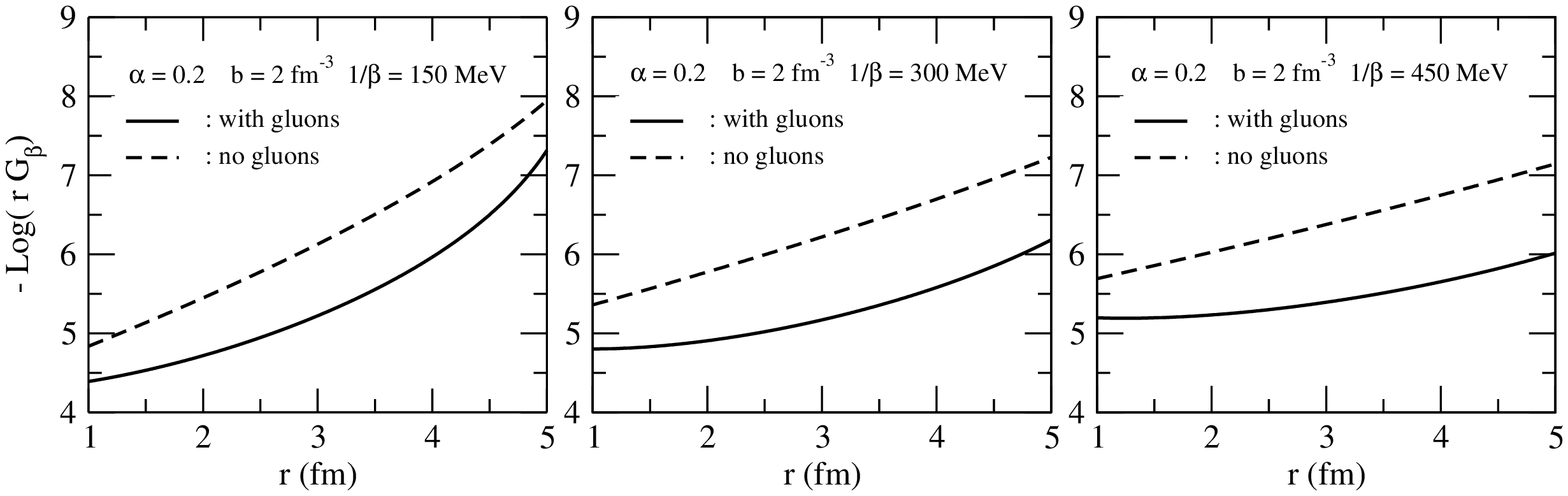}
\label{qq-gg-correlation-lenght}
\caption{Minus Logarithm of $r\,G_{\beta}(r)$  with (solid line) and without (dashed line) curve for different choices of temperature, coupling constant $\alpha=0.2$ and initial quark density $b=2\,fm^{-3}$.}
\vskip0.9cm
\includegraphics[scale=0.40,angle=270]{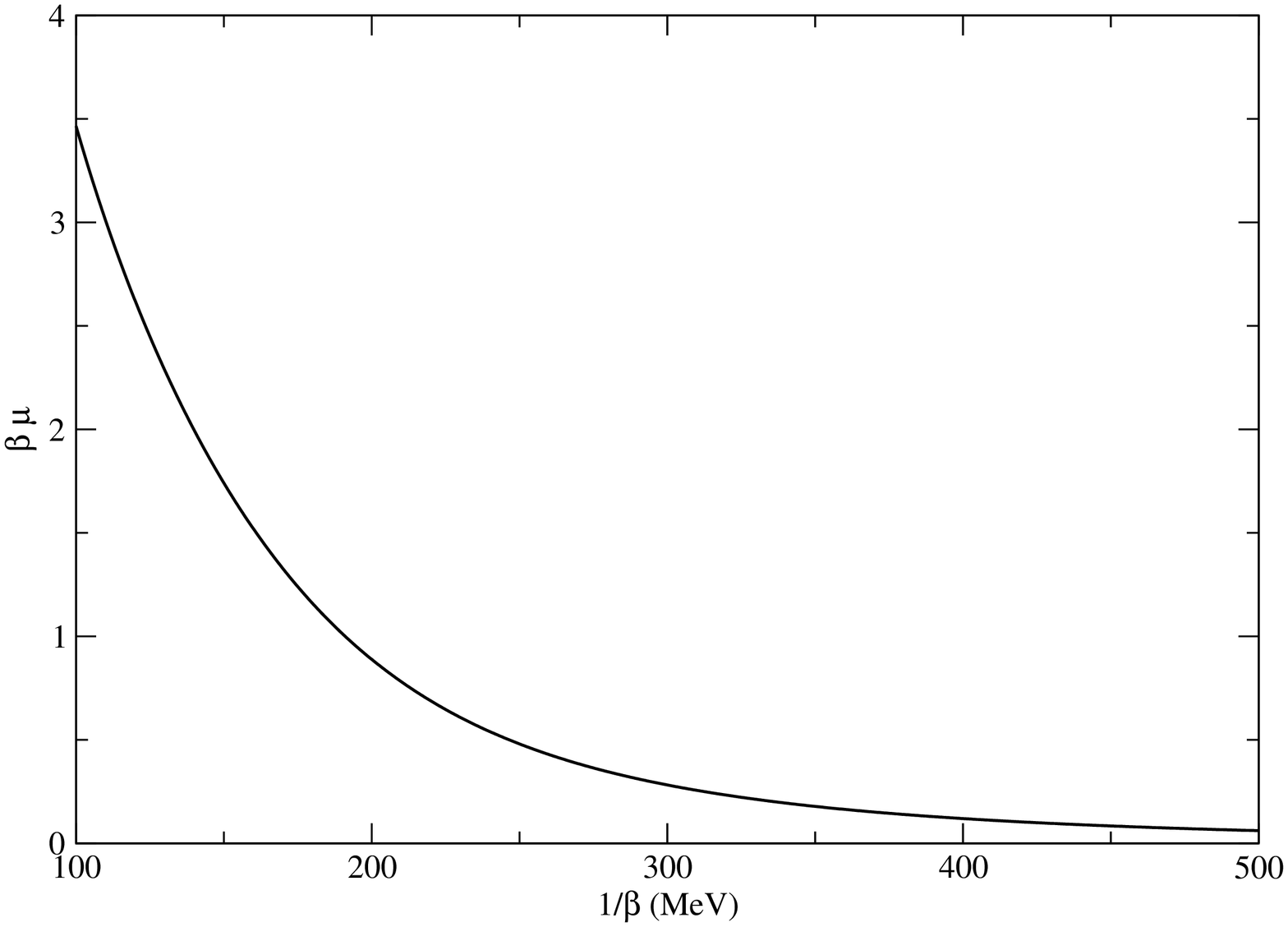}
\caption{$\beta\ \mu$ as a function of $1/\beta$ with initial quark density $b=2\,fm^{-3}$; note that it is an adimensional quantity.} 
\label{betamu}
\vskip1.cm
\includegraphics[scale=0.40,angle=270]{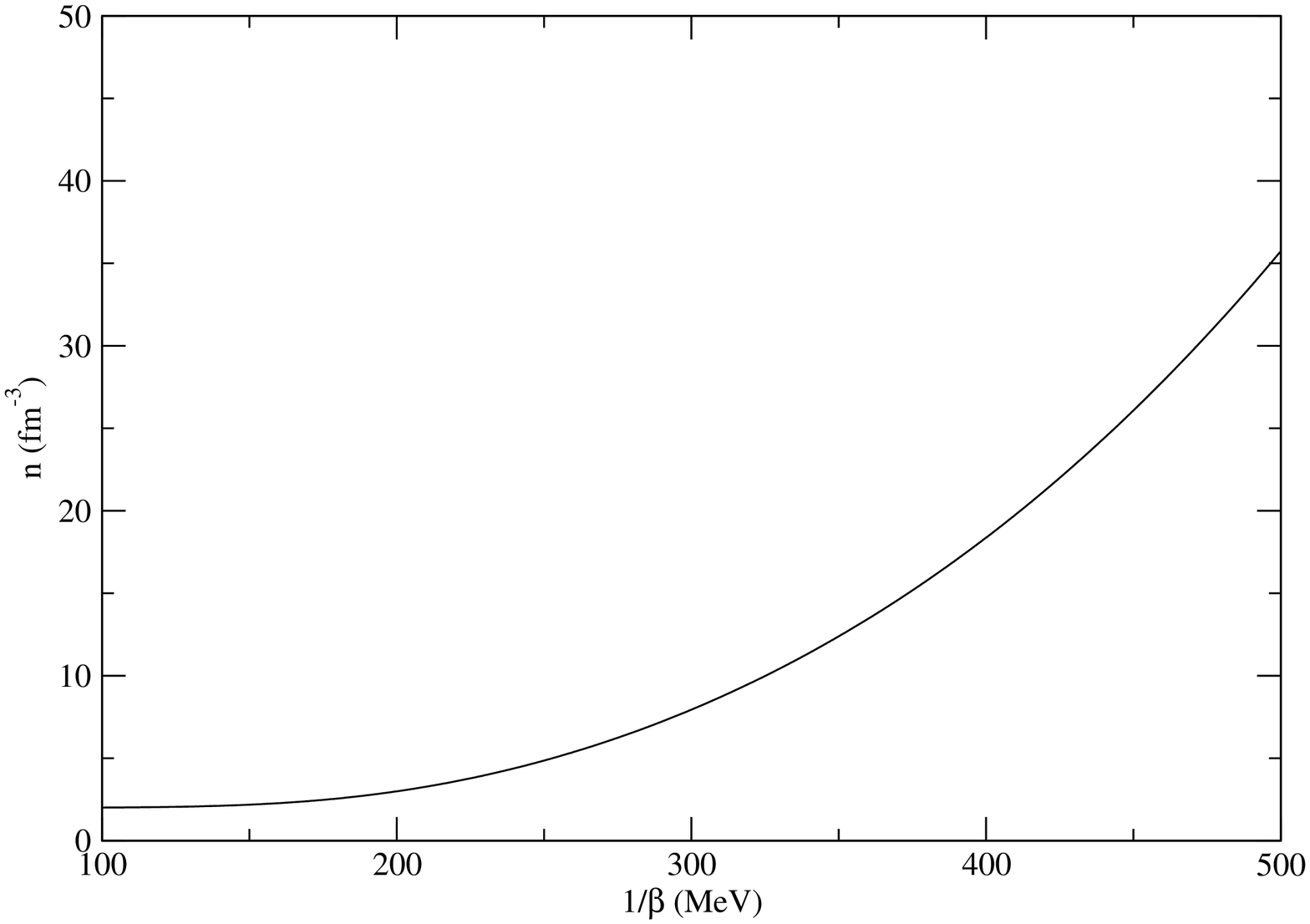}
\caption{Fermion density $n$ as a function of $1/\beta$ with initial quark density given by $b=2\,fm^{-3}$.} 
\label{ndensity}
\end{center}
\end{figure}

It must be noted that the dependence of the shielding effect on the temperature
has two origins: one is the kinematical effect that is present in every
plasma-like system, the other is dynamical and typical of a relativistic system since it  comes from the thermal production
of particles, both gluons and $q\bar q$ pairs. The two processes have not the
same role because initially there are more quarks than
antiquarks, however, looking at the evolution of the chemical potential FIG.\ref{betamu}, we see 
that there is an equivalent temperature, not extremely high, at which the
effects of the initial condition become very small. One could also ask what the
answer would be in the opposite limit, low temperature and high density, which
might be realized inside a neutron star, but in this situation
the Pauli principle would be very relevant and one should start from the
consideration of a degenerate relativistic plasma, what is not attempted in this
paper. 
\begin{center}
\bf{Appendix A: Quark-gluon interaction}
\end{center}
As explained previously only the quark-gluon interaction corresponding to a
gluon exchange in $t-$channel is considered. For this amplitude tha color
structure is usually given in mixed form \cite{Predazzi} as $f_{ABC}(T_C)^a_b$ where
$A,B,C=1\dots 8$ and $a,b=1,2,3$. This form is not convenient here and will be
transformed in a purely spinorial version. Using the fundamental commutator 
$[T_A,T_B]=i f_{ABC}T_C$ we express the interaction as commutator;
then defining the matrices with spinorial indices
$$(T^k_l)^a_b=\frac {1} {\sqrt 2}\big[\delta^a_l \delta^k_b-
\third \delta^a_b \delta^k_l \big] $$
the form of the interactions used in this paper for $qq$ and for $qg$ is

 $$I_{a,c}^{b,d}=
  \frac{1}{2}\left[\delta_a^d\,\delta_c^b-
  \frac{1}{3}\,\delta_a^b\,\delta_c^b\right]\qquad\qquad
  J_{a,cf}^{b,dg}=\frac{1}{2\,i}
  \left[\delta_a^d\,\delta_c^b\,\delta_f^g-
  \,\delta_a^g\,\delta_c^d\,\delta_f^b\right]$$
 Note that the antisymmetry of $J$ with respect to the exchange
 $(c,d)\leftrightarrow (f,g) $ together with the factor $i$
 ensures the Hermiticity of the interaction.
\vskip 1pc 
\begin{center}
\bf{Appendix B: Densities of quarks, antiquarks and gluons}
\end{center}
When we take care of the fact that there can be a thermal production of gluons
we cannot ignore the concurrent production of quark-antiquark pairs. We
investigate briefly the problem in this form. Assume the baryonic density, 
$i.e.$ the quark density minus the antiquark density, as given and find the
simultaneous production of gluons and of quark-antiquark pairs.
The expression for the baryonic density is simple when the quark mass can be 
neglected, or gives at
most a small correction, in this case for the density we get:
\begin{equation}
b=\frac{g_f}{2\,\pi^2}\,\int_0^{\infty}d\epsilon\,\Big(\epsilon^2-\half m^2\Big)
\left[\frac{1}{e^{\beta\,(\epsilon-\mu)}+1}-
\frac{1}{e^{\beta\,(\epsilon+\mu)}+1}\right]
\label{initial-density}
\end{equation}
In particular the antiquark density is given by
\begin{equation}
\bar \rho=\frac{g_f}{2\,\pi^2}\,
\int_0^{\infty}\frac{d\epsilon\,(\epsilon^2-\half m^2)}
{1+e^{\beta\,(\epsilon-\mu)}}
=-\frac{g_f}{2\,\pi^2}\,\Big[\frac{2}{\beta^3}\,
\mathcal{L}_3\left(-e^{-\beta\,\mu}\right)+\frac{m^2}{2\beta}\,
\ln(1+e^{-\beta\,\mu})\Big]
\end{equation}
where $\mathcal{L}_k(z)=\sum_{n=1}^{\infty}{z^n}/{n^k}$ and for the weight we
get $g_f=12$ resulting from a factor 2 for the spin, a factor 3 for the color, a 
factor 2 for the flavor because we consider only $u$ and $d$ quarks.\par
 From eq (\ref{initial-density}) it follows:
\begin{equation}
b=\frac{g_f}{6\,\pi^2}\,\mu^3+\frac{g_f}{6\,\beta^2}\,\mu-\frac{g_f}{4\,\pi^2}
m^2\,\mu \end{equation}
From this relation we see the conditions qualitatively well known in which the
mass term is negligible: either $m<<1/\beta$ or $m<<\mu$, form now on we shall
assume that at least one of these conditions is fulfilled and the mass term is
dropped.
Solving the equation for $\mu$ we find
\begin{equation}
\beta\,\mu=\frac{\left[\left(\frac{27\,\pi^2\,x}{g_f}\right)
\left(1+\sqrt{1+\left(\frac{\pi\,g_f}{9\,\sqrt 3\, x}\right)^2}
\right)\right]^{\frac{2}{3}}-3^{\frac{1}{3}}\,\pi^2}{3^{\frac{2}{3}}
\,\left[\left(\frac{27\,\pi^2\,x}{g_f}\right)
\left(1+\sqrt{1+\left(\frac{\pi\,g_f}{9\,\sqrt 3\, x}\right)^2}
\right)\right]^{\frac{1}{3}}}
\end{equation}
with $x=b\,\beta^3$.
In this way we get the actual expression of $\bar \rho $ in terms of $b$ and
$\beta$, the total density which appears in the previous formulae is given by
$n=b+2\bar \rho $.\par
\begin{figure}[t]
\begin{center}
\includegraphics[scale=0.40,angle=270]{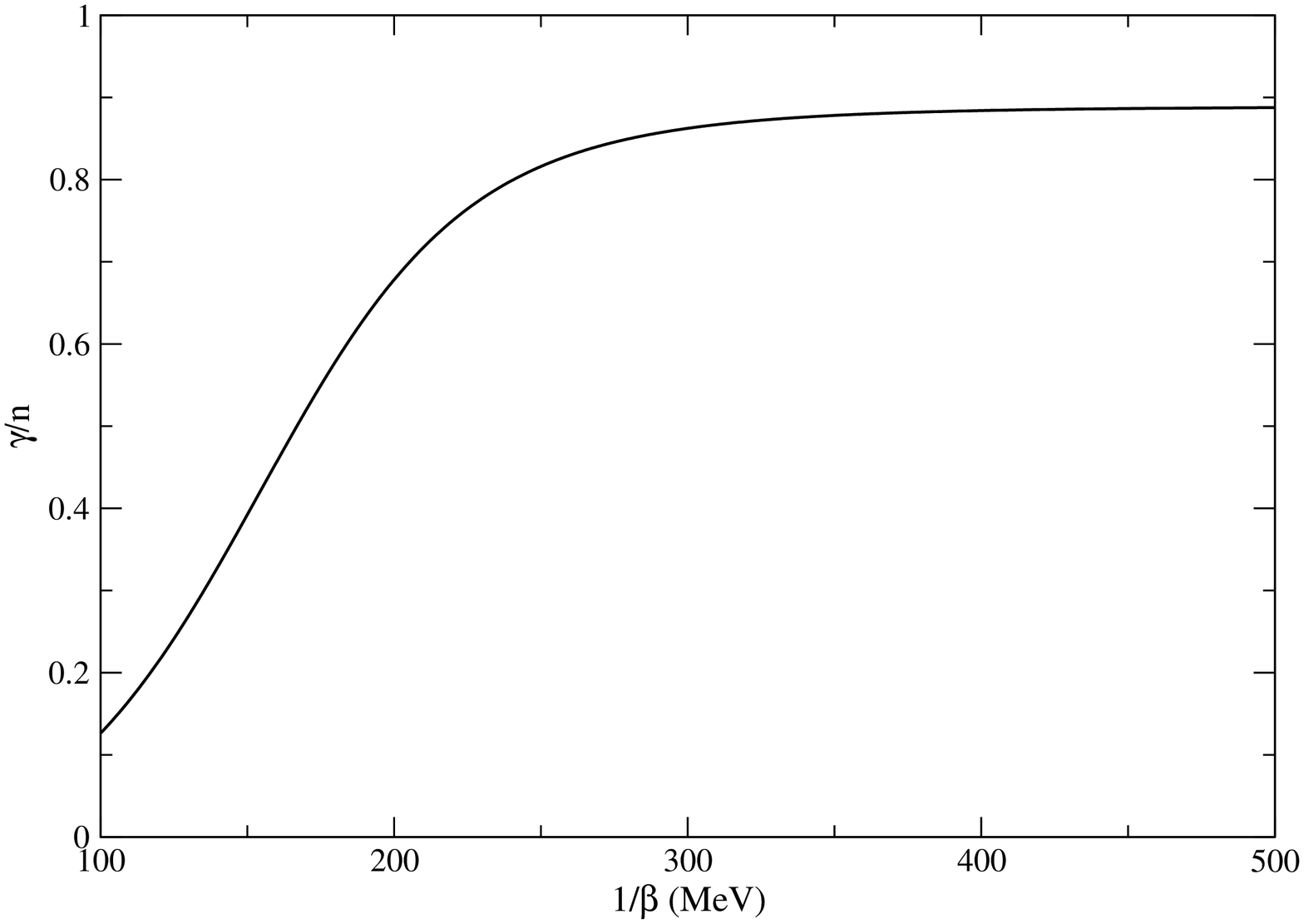}
\caption{$\gamma/n $ ratio as a function of $1/\beta$ with initial quark density given by $b=2\,fm^{-3}$.} 
\label{gammanratio}
\end{center}
\end{figure}

It is now possible to give a semi-quantitative analysis of the different
densities which are relevant for our problem: we start assigning a density $b$
which we take equal to 2 fm${}^{-3}$. 
In FIG.\ref{betamu}, FIG.\ref{ndensity} and FIG.\ref{gammanratio}  the values of $\beta\mu$, $n$ and the ratio $\gamma/n$ as functions of $1/\beta$ are shown.\par
We have worked out three cases: the first is chosen so that the density of antiquarks is much less than $b$, we take $1/\beta=150\,MeV$, this gives $\bar \rho \approx \frac{1}{20}\,b$  and for the corresponding gluon density $\gamma/n = 0.40$, in 
the second case we take $1/\beta=300\,MeV$, this gives $\bar \rho \approx \frac{3}{2}\,b$ and $\gamma/n = 0.84$, in the third one  
$1/ \beta=450\,MeV$ is considered, which produces $\bar \rho \approx 6\,b$ and $\gamma/n = 0.88$, which is the ''saturation'' value for $\gamma/n $ ratio.
 In these expression the density of gluons has be taken as the free-boson
 thermal density, which amounts to:
  \begin{equation}
   \frac {g_b}{2\pi^2} \frac {2} {\beta^3} \mathcal{L}_3 (1)
\end{equation}
 Here $g_b=16\;i.e.$ a factor 2 for the spin and a factor 8 for the color. Since
 we have $\beta\mu\to 0$ for $\beta\to 0$ it results that in that limit
  $$ \frac {\gamma} {n}=\frac {g_b} {2g_f} \frac {\mathcal{L}_3 (1)}
  {-\mathcal{L}_3 (-1)}=\frac {8}{9}$$
  and this limit is numerically reached already at $1/\beta=390$ MeV.

\vskip 1pc 
\begin{center}
\bf{Appendix C: Analytical expression of the correlations}
\end{center}
 The inversion of the Fourier transform
\begin{equation}
G_{\beta}(r^2)=\frac{1}{(2\,\pi)^3}\int d^3k\,e^{i\,\mathbf{k \cdot r}}\,
\hat G_{\beta}(k^2)\nonumber
\end{equation}
 implies a standard angular integration and then an integral over the radial
 coordinates that gives
\begin{align}
G_{\beta}(r^2)=\frac{1}{2\,\pi\,r^2}&\left[g_{\beta}^1\,\,
{}_0F_2\left(;\frac{1}{2},\,2;\,\frac{3\,\pi\,n\,\alpha\,\beta\,(1-i\,y)
}{4}\,r^2\right)+g_{\beta}^2\,\,{}_0F_2\left(;\frac{1}{2},\,2;\,
\frac{3\,\pi\,n\,\alpha\,\beta\,(1+i\,y)}{4}\,r^2\right)\right.\nonumber \\
+&\left. g_{\beta}^3\,\,{}_0F_2\left(;\frac{3}{2},\,\frac{5}{2};\,
\frac{3\,\pi\,n\,\alpha\,\beta\,(1-i\,y)}{4}\,r^2\right)+
g_{\beta}^4\,\,{}_0F_2\left(;\frac{3}{2},\,\frac{5}{2};\,
\frac{3\,\pi\,n\,\alpha\,\beta\,(1+i\,y)}{4}\,r^2\right)  \right]
\end{align}
where the generalized hypergeometric functions are used
\begin{equation}
{}_pF_q\left(a_1,\ldots,a_p\,;\,b_1,\ldots,b_q\,;z\right)=
\sum_{k=0}^\infty\, \frac{(a_1)_k\cdots(a_p)_k}{(b_1)_k\cdots(b_q)_k}\,\frac{z^k}{k!}
\end{equation}
and the coefficient are
\begin{align}
g_{\beta}^1&=\frac{\pi^2\,\alpha\,\beta\,(i+y)}{2\,y}\left(1+\frac{6\,\gamma/n}{1+6\,\gamma/n}\right)\,\left(
\frac{1-12\,\gamma /n}{3\,(1+12\,\gamma /n)}-i\,y\right)\nonumber\\
g_{\beta}^2&=\frac{\pi^2\,\alpha\,\beta\,(-i+y)}{2\,y}\left(1+\frac{6\,\gamma/n}{1+6\,\gamma/n}\right)\,\left(
\frac{1-12\,\gamma /n}{3\,(1+12\,\gamma /n)}+i\,y\right)\nonumber\\
g_{\beta}^3&=\frac{\sqrt{3\,\pi\,n\,(4\,\pi\,\alpha\,\beta)^3}\,(1-i\,y)^{\frac{3}{2}}\,r}{12\,(i\,y)}\left(1+\frac{6\,\gamma/n}{1+6\,\gamma/n}\right)\,\left(
\frac{1-12\,\gamma /n}{3\,(1+12\,\gamma /n)}-i\,y\right)\nonumber\\
g_{\beta}^4&=\frac{\sqrt{3\,\pi\,n\,(4\,\pi\,\alpha\,\beta)^3}\,(1+i\,y)^{\frac{3}{2}}\,r}{12\,(-i\,y)}\left(1+\frac{6\,\gamma/n}{1+6\,\gamma/n}\right)\,\left(
\frac{1-12\,\gamma /n}{3\,(1+12\,\gamma /n)}+i\,y\right)\nonumber.\\
\end{align}
The function $G_{\beta}(r^2)$ is real because it is the sum of two terms with their complex conjugate. Note that for very low values of $1/\beta$, $y$ could become imaginary, in this case all the addenda would be separately real.    


\begin{thebibliography}{99}
\bibitem{Calucci}
G.~Calucci, Eur. Phys. J. C \textbf{36}, 221-226 (204).
\bibitem{Birse}
M.C.~Birse, C.W.~Kao, G.C.~Nayak, Screening and antiscreening in an anisotropic QED and QCD plasma [hep-ph/0304209]; O.~Philipsen, What mediates the longest correlation length in the QCD plasma? [hep-ph/0301128].
\bibitem{Philipsen}
O.~Philipsen,~Phys.~Lett. B \textbf{521}, 273 (2001).
\bibitem{Kogut}
J.~Kogut, L.~Susskind, Phys. Rev. D \textbf{11}, 395 (1975).
\bibitem{Le-Bellac}
M.~Le~Bellac, Thermal Field Theory, (Cambridge U.P., Cambridge (1996)).
\bibitem{Predazzi}
E.~Leader, E.~Predazzi, An introduction to gauge theories and modern particle physics (Cambridge U.P., Cambridge 1996), App.2.
\bibitem{Vegas}
G.~P.~Lepage, J. Comp. Phys. \textbf{27}, 192 (1978).
\end{thebibliography}
\end{document}